\documentclass[conference]{IEEEtran}
\IEEEoverridecommandlockouts
\usepackage{cite}
\usepackage{amsmath,amssymb,amsfonts}
\usepackage{algorithm}
\usepackage{algpseudocode}
\usepackage{hyperref}
\usepackage{graphicx}
\usepackage{textcomp}
\usepackage{xcolor}
\usepackage[acronym,toc]{glossaries}
\newacronym{hvac}{HVAC}{Heating, Ventilation and Air Conditioning}
\newacronym{pmv}{PMV}{Predicted Mean Vote}
\newacronym{tci}{TCI}{Thermal Comfort Index}
\newacronym{gsr}{GSR}{Galvanic Skin Response}
\newacronym{ble}{BLE}{Bluetooth Low Energy}
\newacronym{hse}{HSE}{Health and Safety Executive}
\newacronym{hr}{HR}{Heart Rate}
\usepackage{url}
\usepackage{soul}
\usepackage{comment}
\usepackage{orcidlink}

\usepackage{url}
\usepackage{amsmath}
\usepackage{courier}
\usepackage[T1]{fontenc}

\usepackage{lipsum}
\usepackage{graphicx}

\def\BibTeX{{\rm B\kern-.05em{\sc i\kern-.025em b}\kern-.08em
    T\kern-.1667em\lower.7ex\hbox{E}\kern-.125emX}}
\begin{document}

\title{Towards Achieving Thermal Comfort through Physiologically Cloud based controlled HVAC System\\

\thanks{$^{1}$ The authors are with the Department of Computer Science, School of Science and Technology, Nottingham Trent University, Clifton Lane,
NG11 8NS, United Kingdom \texttt{\{isibor.ihianle, pedro.machado, kayode.owa, david.adama\}@ntu.ac.uk}
}
}

\author{\IEEEauthorblockN{Isibor Kennedy Ihianle$^{1}$\orcidlink{0000-0001-7445-8573}, Pedro Machado$^{1}$\orcidlink{0000-0003-1760-3871}, Kayode Owa$^{1}$\orcidlink{0000-0002-1393-705X} and David Ada Adama$^{1}$\orcidlink{0000-0002-2650-857X}}
}

\maketitle

\begin{abstract}
Thermal comfort in shared spaces is essential to occupants’ well-being and necessary in the management of energy consumption. Existing thermal control systems for indoor shared spaces adjust temperature set points mechanically, making it difficult to intelligently achieve thermal comfort for all. Recent studies have shown that thermal comfort in a shared space is difficult to achieve due to individual preferences and the inability of occupants to reach a thermal compromise on temperature set points. This paper proposes a thermal comfort system to automatically adjust the temperature set-points in a shared space whilst recognising individual preferences. The control strategy of the proposed system is based on an algorithm to adjust the temperature set point of the shared space using the individual thermal preferences and predicted thermal comfort value of the occupants. The thermal preferences of the occupants are determined first and used as part of the occupant’s profile, which is mapped to thermal comfort values predicted from the occupants’ measured physiological data and environmental data. A consensus is reached by the algorithm to find the optimal temperature set-point, which takes into account individual thermal preferences and their physiological responses.
\end{abstract}


\section{Introduction}
Recent advances in the use of sensing devices in different application domains are promising and offer huge possibilities given their sizes, low cost, and capabilities. One of such areas of application is in smart home systems to predict the thermal comfort of individuals in their homes \cite{ref1, ref2, ref3}. The provision of thermal comfort in home environments, offices, and shared spaces can increase activity, productivity, and overall well-being of the occupants. Alajmi et. al. \cite{ref44} and Langevin et. al. \cite{ref444} identified thermal comfort as the most important factor affecting occupants' productivity in shared spaces. Typically, thermal comfort provisions in most buildings are limited to fixed temperature settings, which do not include the occupants in the control loop. These temperature settings ignore the peculiar nature and requirements of human comfort, thus violating the \gls*{hse} recommendations for a reasonable comfort environment \cite{ref11}. According to a recent study, 40\% of the surveyed buildings were colder than the recommended air temperature, and 61\% of occupants felt too cold, and this over-cooling resulted in an estimated annual waste of 18.7 million kWh during the summer months \cite{ref10}. In cases where occupants have access to the thermal control, there is a lack of compromise and negotiation. A recent study, found that in a multi-occupancy setting, the occupants were oblivious to others' feelings and the need to compromise on a temperature set point \cite{ref5}. This lack of compromise and the ignorance of the idiosyncratic nature of the occupants makes it hard for thermal comfort provisions and, in some cases, drives it to extremes \cite{ref4}. The inability to provide shared spaces that are responsive to the peculiar nature and requirements of the occupants becomes a key challenge. Although, it is not easy to provide occupants with thermal comfort, by focusing on the physical variables and the physiological response of the occupants, thermal comfort can be encouraged and enhanced.

\begin{figure*}
    \centering
    \includegraphics[width=15cm,height=7cm]{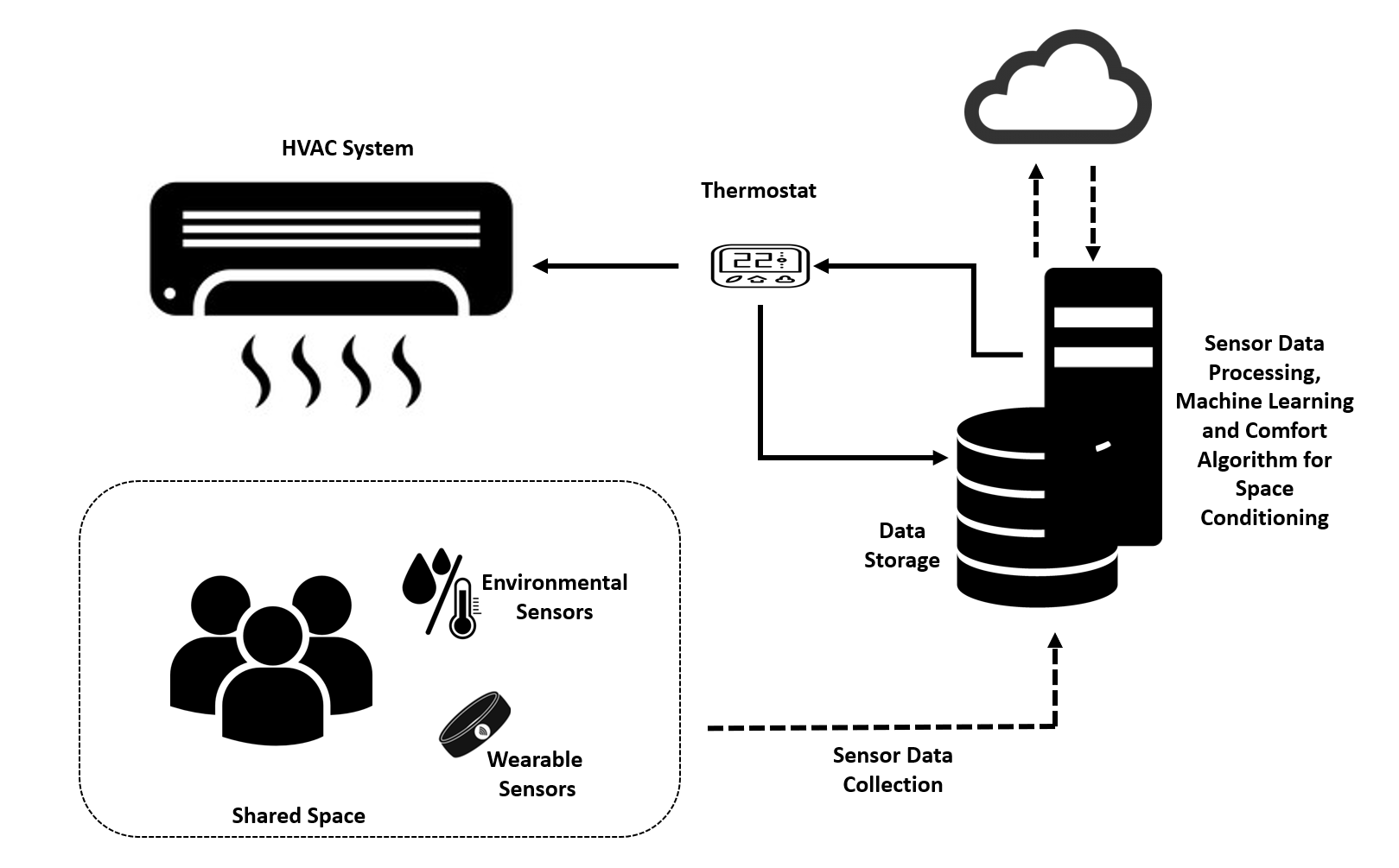}
    \caption{Overview of the proposed thermal comfort control System}
    \label{fig:img1}
\end{figure*}

To overcome this challenge, this paper proposes a system to estimate thermal comfort  of the individuals in a shared space by measuring physiological responses as the relationship between the occupants' thermal satisfaction level to accurately estimate the thermal satisfaction of the occupants in the shared space. Previous studies have linked thermal sensation as encompassing the physiological and subjective response of the occupants to the thermal environment and the shared space they occupy\cite{ref1, ref88, ref888}. The aim of the proposed approach is to solve this thermal preference compromise and negotiation problem through learning individual thermal profiles. The average of each occupant's desired temperature is used, which is the  current common practice. The temperature range that some people can tolerate varies from that of others. By allowing the less sensitive occupant to accept greater compromises, this approach would have the potential of achieving thermal comfort for all. The percentage of thermal comfort could be increased if we are able to get such learned individual thermal profiles. The remainder of the paper is organized as follows: Section \ref{sec:lr} provides an overview of the related works, Section \ref{sec:proposed_sys} describes the proposed approach for thermal comfort and Section \ref{sec:alg} thermal comfort algorithm, while Section \ref{sec:lim} highlights the challenges and limitations. Finally, conclusions and future work are discussed in Section \ref{sec:conclusions}.

\section{Related Work} \label{sec:lr}

The provision of thermal comfort for individuals in buildings is a necessity. In most cases, the room temperatures are set to thermal conditions that are not conducive to the occupants, leading to thermal discomfort. Although these buildings may have temperature models which allow for adjustments, studies \cite{ref1, ref5} have shown that occupants do not have enough knowledge of the \gls*{hvac} control systems and thermostats to change the temperature settings. In multi-occupancy settings, occupants are oblivious of others' feelings and the necessity to compromise with others to achieve a thermally comfortable setting \cite{ref5}. Most thermal comfort provision mechanisms work on the assumption that everyone, regardless of gender, age, or physical or psychological state, has the same expectations for thermal comfort \cite{ref99, ref999}.

The \gls*{pmv} and adaptive model, which are the most widely used thermal comfort model, have been shown to be less effective \cite{ref6}\cite{ref7}. These models ignore the influence of individual differences in physiological, psychological, and behavioural aspects. Recent studies have shown that physiological measures of an individual exhibit remarkably distinct patterns when they are exposed to a thermally uncomfortable environment from a comfortable setting \cite{ref8, ref12}. Individuals experience thermal sensations as a response to the temperature of their environment, which can be correlated to their physiological and psychological responses. As a result, there is a correlation between physiological parameters and the thermal experience of the occupants, which can be taken advantage of to control systems for thermal comfort.

Physiological measures like skin temperature, sweat rate, heart rate, blood pressure, and brain waves were the most used for the evaluation of the thermal experience\cite{ref8, ref12}. The emergence of sensing technologies provides the opportunity for the inference of thermal comfort from measured physiological data. Numerous studies have been conducted using different sensors and computational models to predict thermal comfort \cite{ref9}.

These individualised thermal comfort models performed satisfactorily well in predicting occupants' thermal experiences. These demonstrate that it is possible to predict an individual’s thermal comfort from the patterns of their physiological signals in real-time using machine learning algorithms. In the context of multi-occupancy, where compromise is a challenge, it is necessary to automate control systems to create a smart system that is responsive to the collective physiological patterns of the occupants.

\section{The Proposed Thermal Comfort System} \label{sec:proposed_sys}
The thermal conditions of most buildings are typically set to temperature points where occupants desire in order to re-adjust in response to temporary changes in hot or cold situations, resulting in increased discomfort. These adjustments should ideally be automated to satisfy the occupant whilst taking into consideration the conflicting comfort preferences of other people in the same space to create a more responsive shared space. Several studies on the use of physiological data in the development of personal thermal comfort models have been conducted \cite{ref1, ref6, ref8}. These make it possible for thermal comfort models to recognise individual preferences by taking advantage of their physiological data in a multi-occupancy setting. As mentioned earlier, this paper proposes an approach to solve this thermal preference compromise and negotiation problem through learning individual thermal profiles. Fig.~\ref{fig:img1} illustrates the schematic of the proposed thermal comfort solution.

\begin{figure*}
    \centering
    \includegraphics[width=13cm,height=6cm]{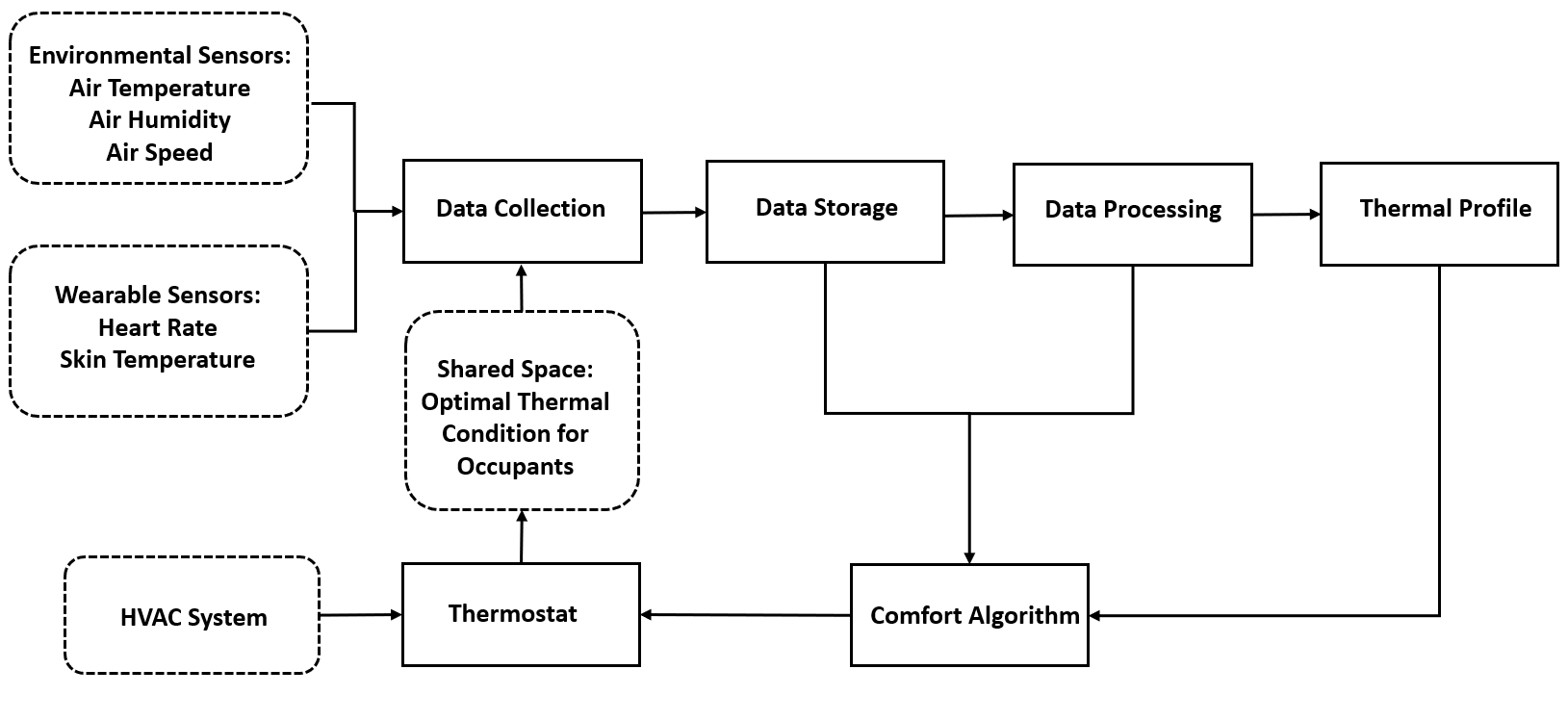}
    \caption{Schematic of Working principle for the Data gathering and processing}
    \label{fig:img2}
\end{figure*}

The system's three main components are depicted in the diagram: data gathering via wearable devices and indoor ambient thermal condition sensors; thermal comfort modelling and intelligent control module; and the HVAC system to provide space conditioning. The data gathering module is made up of wearable sensors and ambient sensors to collect environmental data such as room temperature and humidity. It is envisaged that the sensors use \gls*{ble} connection to interact with the thermal comfort modelling and intelligent control module. The thermal comfort modelling and intelligent control module are cloud-based and act as the main hub and server to provide services to other sub-components with data storage and processing capabilities.

Fig.~\ref{fig:img2} summarises the proposed data collection and data flow. Environment sensors are installed in the building and smart wearable bands are worn by the occupants for data collection, each of which is considered as a node with a unique identifier for cloud-based thermal comfort modelling and intelligent control module. Data captured by the sensors is analysed so that the relationship between the occupants' actual thermal comfort indices and their thermal profiles are established through a thermal comfort algorithm. The system then adjusts and regulates the temperature inside. The services of the thermal comfort and intelligent control modules are integrated into the controls to determine the adaptive set-point based on the occupants' thermal profile.

\section{Thermal Comfort Algorithm} \label{sec:alg}
This research is aimed at the provision of a thermally conducive space for occupants whilst taking into consideration conflicting individual thermal preferences. Modelling preferences for thermal comfort can be achieved using machine learning algorithms on the occupants’ physiological data and environmental variables to determine the \gls*{tci}. Similar to Fanger’s \gls*{pmv} model, the \gls*{tci} can be expressed from -3 to 3 with 0 as thermally comfortable or neutral \cite{ref6}. The \gls*{pmv} is determined from air temperature, air humidity, clothing insulation, air velocity, metabolic rate, and mean radiant temperature. Studies have also shown that \gls*{hr} can be used as a predictive biomarker for thermal comfort \cite{ref8, ref13}. In this work, we propose using machine learning algorithms to predict \gls*{tci} from physiological data like \gls*{hr}, \gls*{gsr} features extracted using wearable smart bands, clothing insulation in addition to environmental variables like air temperature, air velocity, and air humidity. It is envisaged that with occupants in a shared space, physiological data can be collected using wearable smart bands as part of the thermal comfort strategy. The thermal comfort strategy to adjust the  thermostat is defined in Algorithm 1.  First, a group thermal comfort profile is developed using machine learning to model the group’s thermal comfort indices from the physiological data of all the possible occupants and the environmental factors. The optimal temperature set point $T_0$ which considers the sensitivity of all the occupants, will be based on the spread of the air temperature readings as ruled by equation \ref{eq:1}.
\\
\begin{equation} \label{eq:1}
 T_0 = \bar{T} \pm \sqrt{\frac{1}{T}\sum_{i=1}^T(T_i-\bar{T})^2}
\end{equation}
\\

where $T$ = {$t_1..., t_n$} and $\bar{T}$ are temperature values of all the occupants and the average respectively for which the group’s thermal comfort index is 0 or neutral.

Thermal adjustment for comfort can be achieved by algorithm \ref{alg:1} correlating predicted \gls*{tci} to the group thermal comfort profile for the optimal air temperature since it is the only parameter that can be controlled. For example, if the occupants in a shared space have predicted \gls*{tci} less or greater than the thermal neutral value of 0, then algorithm \ref{alg:1} adjusts the thermostat so that air temperature is equivalent to the corresponding air temperature value where the group thermal profile points to a value of 0 or neutral as the new temperature set point for thermal comfort. This process can be repeated as needed to allow the system to adjust to changes in the shared space.

\algnewcommand\algorithmicforeach{\textbf{for each:}}
\algnewcommand\ForEach{\item[ \algorithmicforeach]}

\begin{algorithm} \caption{Thermal Comfort Algorithm} \label{alg:1}
\begin{algorithmic}[1]
\Require $Group Thermal Profile, T_0$
\State $HVAC \gets ON$
\State $Wearable\_Sensors \gets ON$
\State $Env\_Sensors \gets ON$
\Repeat
\ForEach {$Occupants  \geq 1$};
\State $Compute\_TCI\_of\_occupants$
\If {$0 < TCI > 0$} 
        \State $New\_Thermostat\_Set\_Point \gets T_0$ 
    \EndIf
    
\Until $Air\_temp \approx T_0$
\end{algorithmic}
\end{algorithm}

\section{Challenges and Limitations} \label{sec:lim}

Although the proposed approach offers opportunities to provide thermal comfort in shared spaces, it also comes with potential challenges and limitations worth considering as part of its implementation.

\subsection {Connectivity and Power Issues}
Connectivity and coverage of the sensor nodes pose a major challenge and remain important concerns for the proposed approach. The sensor nodes rely on good \gls*{ble} and internet connectivity to interact with the cloud-based thermal comfort modelling and intelligent control module. Poor coverage and connectivity can significantly impact the performance of the sensor nodes, thereby negatively impacting the data collection and storage process and the reliability of the thermal comfort algorithm. However, issues relating to data storage can be addressed using hybrid solutions of local data collection and multi-region storage mechanisms.

Sensor power consumption poses another challenge. The sensors have limited energy sources, so proper activation of the sensors can have a significant influence on the data collection. An on-board processing sensor node consumes power and stops data collection as soon as it runs out of power. A solution to this could be to process the data in a cloud-based thermal comfort modelling and intelligent control module. However, data security must be guaranteed.

\subsection {Privacy and Security Issues} 
Buildings and shared spaces intended to adapt to the proposed system must be designed to protect data collected from data breaches, unauthorised access and external threats. To limit access and intrusion and maintain data integrity, data privacy and cyber-security must be integrated into the building and shared space as a functionality.

\section{Conclusions and Future Work} \label{sec:conclusions}
This paper proposes a thermal comfort system to automatically adjust the temperature set-points in a shared space whilst recognising individual preferences. The thermal comfort algorithm finds the optimal temperature set-point which considers individual thermal preferences and their predicted physiological responses given environmental factors. The occupants’ thermal preferences are determined first and then mapped to the predicted thermal comfort values based on the occupants’ measured physiological data and environmental data. Smart wearable devices are used to capture the occupant’s physiological data and other sensors to collect environmental data. The proposed thermal comfort system goes beyond the current approaches to include a centralised cloud-based infrastructure with data storage, data aggregation, and a thermal comfort algorithm to automatically adjust settings. Future plans will be directed to the design and implementation of the proposed approach as a proof of concept.


\section*{Acknowledgment}
This research has been supported by the Nottingham Trent University.


\begin{thebibliography}{00}

\bibitem{ref1} Energy savings and performance gains in gsa buildings, https://www.gsa.gov/cdnstatic/GSA\_SevenStrategies\_090327screen.pdf

\bibitem{ref2} F. Alsaleem, M. Tesfay, M. Rafaie, K. Sinkar, D. Besarla, P. Arunasalam, An iot framework for modeling and controlling thermal comfort in buildings. Frontiers
in Built Environment 6(87) (2020).

\bibitem{ref3} T. Chaudhuri, D.  Zhai, Y. Soh, H. Li, L. Xie, Random forest based thermal comfort prediction from gender-specific physiological parameters using wearable
sensing technology. Energy and Buildings 166, 391 – 406 (2018).

\bibitem{ref44} A. F. Alajmi, F. A. Baddar, R. I. Bourisli, Thermal comfort assessment of an office building served by under-floor air distribution (UFAD) system – A case study. Building and Environment, 85, 153-159 (2015).

\bibitem{ref444} J. Langevin, J. Wen, P. L. Gurian, Modeling thermal comfort holistically: Bayesian estimation of thermal sensation, acceptability, and preference distributions for office building occupants. Building and Environment, 69, 206-226 (2013).

\bibitem{ref4} J. H. Choi, D. Yeom, Development of the data-driven thermal satisfaction prediction model as a function of human physiological responses in a built environment.
Building and Environment 150, 206–218 (2019).

\bibitem{ref5} B. Dong, V. Prakash, F. Feng, Z. O’Neill, A review of smart building sensing system for better indoor environment control. Energy and Buildings 199, 29–46
(2019).

\bibitem{ref6} F. O. Fanger, Thermal comfort. analysis and applications in environmental engineering, thermal comfort. Analysis and Applications in Environmental Engineering, Danish Technical Press (1970).

\bibitem{ref7} HSE: Thermal comfort. https://www.hse.gov.uk/temperature/thermal/ (2022).

\bibitem{ref8} C. Huang, R. Yang, M. Newman, The potential and challenges of inferring thermal comfort at home using commodity sensors. In: Proceedings of the 2015 ACM
International Joint Conference on Pervasive and Ubiquitous Computing. Osaka,
Japan, (2015).

\bibitem{ref88} J. H. Choi and V. Loftness, Investigation of human body skin temperatures as a biosignal to indicate overall thermal sensations, Build. Environ., vol. 58, no. null, pp. 258–269 (2012).

\bibitem{ref888} M. Indraganti, K. D. Rao, Effect of age, gender, economic group and tenure on thermal comfort: A field study in residential buildings in hot and dry climate with seasonal variations, Energy Build., vol. 42, no. 3, pp. 273–281 (2010).

\bibitem{ref99} S. Karjalainen, Gender differences in thermal comfort and use of thermostats in everyday thermal environments, Build. Environ., vol. 42, no. 4, pp. 1594–1603, (2007).
\bibitem{ref999} S. Karjalainen, Thermal comfort and gender: A literature review, Indoor Air, vol. 22, no. 2, pp. 96–109, (2012).

\bibitem{ref9} M. Humphreys, F. Nicol, The validity of iso-pmv for predicting comfort votes in every-day. Energy and Buildings 34, 667 – 684 (2002).

\bibitem{ref10} J. Kim, Y. Zhou, S. Schiavon, P. Raftery, G. Brager, Personal comfort models: Predicting individuals’ thermal preference using occupant heating and cooling behavior and machine learning. Buildings and Environment 129, 96–106 (2018).

\bibitem{ref11} S. Liu, S. Schiavon, H. Das, M. Jin, C. Spanos, Personal thermal comfort models with wearable sensors. Building and Environment 162, 106281 (2018).

\bibitem{ref12} K. Nkurikiyeyezu, Y. Suzuki, G. Lopez, Heart rate variability as a predictive biomarker of thermal comfort. Journal of Ambient Intelligence and Humanized
Computing 9, 1465–1477 (2018).
 
\bibitem{ref13} D. Zhipeng, C. Qingyan, Artificial neural network models using thermal sensations and occupants’ behavior for predicting thermal comfort. Energy and Buildings 174, 587–602 (2018).


\end{thebibliography}
\end{document}